\documentclass[12pt]{article}
\usepackage{amsmath,amssymb}
\usepackage{braket}
\usepackage{graphicx}
\usepackage{xcolor}
\usepackage{dcolumn}
\usepackage{bm}
\usepackage[numbers,super,comma,sort&compress]{natbib}

\usepackage{authblk}
\definecolor{background-color}{gray}{0.98}

\title{Quantum Monte Carlo method for metal catalysis: case study of hydrogen production on Pt(111)}

\author{Rajesh O. Sharma$^1$, Tapio T Rantala$^2$ and Philip E Hoggan*$^1$}

\affil{1 Institute Pascal, UMR 6602 CNRS,BP 80026, 63177 Aubiere Cedex, France, \\ 2 Physics, Tampere University, Tampere, Finland. }

\date{}

\begin{document}

\maketitle
\footnote{email for correspondence: philip.hoggan@uca.fr}
\begin{abstract}

  Over 90 \% of all chemical manufacture uses a solid catalyst. Related work thus responds to major societal demand. This study is of water-gas shift on platinum for hydrogen production. The close-packed Pt(111) surface catalyses this process.

Many chemical reactions involve bond-dissociation. This is also true for reactions at solid surfaces, in which the dissociation step is often limiting but facilitated in comparison to gas phase reaction channels. Since bond-breaking is poorly described by Hartree-Fock and DFT methods, this work adopts Quantum Monte Carlo (QMC) methodology. QMC is a stochastic approach to solving the Schr{\"o}dinger equation recently came of age for heterogeneous systems involving solids.

The present work considers co-adsorption of water and carbon monoxide on Pt(111). The water is partially dissociated while its oxygen atom binds to CO losing a hydrogen atom. This concerted step is rate-limiting. The resulting adsorbed formate species then decomposes to readily eliminated carbon dioxide and the clean-fuel product is H$_2$.

The Transition-State geometry can be optimized using molecular Quantum Monte Carlo force constants, on the basis of our earlier work using the CASINO software.

Our embedded active site approach is used. This allows a high-level configuration interaction (CI) wave-function to be used, expanded in plane-waves and embedded in the metal lattice exposing its close-packed face. The resulting periodic function is used to guide the Quantum Monte Carlo calculation.

Results are given here on mechanism and QMC activation barrier for water addition to CO pre-adsorbed on Pt(111) of 17 $\pm$ 0.2 c.f. apparent measured value of 17.05 kcal/mol \cite{shar}. They are encouraging for investigating similar or complex catalytic systems.

\end{abstract}

\vskip4mm
\section{Introduction}

\hskip5mm There are few innovative approaches to accurate determination of electronic structure calculations. Nevertheless, several groups have begun to work on the challenges of real systems which require electron correlation accurately.
One such growth area is heterogeneous catalysis, where quantum chemistry meets solid state physics. The former discipline has not completely abandoned atomic orbitals to define basis functions and the latter clings to plane-waves in reciprocal space. The former uses localised co-ordinate space functions difficult to represent in plane waves and the latter, plane waves, delocalised in co-ordinate space but localised in reciprocal space (and easily transformed by Fourier series to co-ordinate space). Of course, there is electron density as a property-related function, although direct access may be difficult. The electron density is described by the Quantum Monte Carlo method used here.
This study focusses on the early stages of the water-gas shift reaction (wgs) which follows the overall equation:
\vskip1mm
CO + H$_2$O $\rightarrow $  CO$_2$ + H$_2$
\vskip4mm
The reaction was discovered in 1780 by Fontana. Much of the mechanistic work has postulated a redox mechanism, particularly for metals like copper which are known to be reducing agents \cite{mavr}. The alternative mechanism is associative and this work shows no change in the Pt-oxidation state whilst proposing an additive complex between CO and water, prior to hydrogen (and CO$_2$) production from formate decomposition. A detailed study of this mechanism on Pt, including experiment, DFT and kinetics is proposed in a recent review \cite{mukin}.
This reference also collects previously published activation barriers for the CO oxidation involved.
These barriers cover a factor of two: ranging from 11.3 to 23.3 kcal/mol  \cite{mukin}.
\vskip4mm
Hence, a key step of water-gas shift reaction mechanisms shows that mainstream Quantum chemistry gives approximate activation barriers for this system. This implies {\it ab initio} Hartree-Fock (HF) (with/without post-HF correlation) or DFT do not deliver meaningful accuracy on these activation barriers.

The present theoretical study of hydrogen production by catalytic addition of water to carbon monoxide adopting the Quantum Monte Carlo (QMC) method therefore represents a breakthrough in reliable information on bond-dissociation limitation of metal catalysed reactions. QMC was bench-marked for hydrogen dissociation on copper, for which accurate molecular beam measurements are available \cite{kdd2}.

Our methodological advances are described here in detail. The main results of an enabling 52Mh supercomputer allocation are alwo given in this work.

A data set collection of computational results is available in the Figshare repository and can be accessed via the repository-generated DOI 10.6084/m9.figshare.10293194 \cite{bcs}.

This work is among the largest heterogeneous systems studied by QMC.
The water-gas shift reaction is greatly facilitated on a platinum catalyst. Reactants are in equilibrium with products and we focussed on theoretical study of the forward reaction in this work, producing hydrogen. We determine the rate-limiting step of the reaction mechanism and the associated activation barrier, on Pt(111).

\vskip4mm

Platinum is a 5d$^9$ 6s$^1$ ground state which can be represented by 10 (or, better, by 18) valence electrons outside a suitable pseudo-potential, noting that little core-valence occupation occurs in any given region of space, because the related densities are well-separated. This confers low variance on the metal wave-function, an order of magnitude lower than that of the (problematic) copper metal wave-functions. A study of the performance obtained with pseudo-potentials discussed below for excitation energies can be found in \cite{Raj1}.



Basis and pseudo-potential choice must be carefully made for the present work. Plane waves with complex exponent allow basis sets to include Stater type orbitals, which facilitates embedding \cite{grun, hog1}.  The requisite method of choice must scale well with system size, be able to efficiently use modern supercomputer facilities that are massively parallel and, above all, produce quantitative, accurate physical properties that are difficult to obtain otherwise. Quantum Monte Carlo benchmarks of activation barriers for reactions adsorbed on solid catalyst surfaces certainly fit this description.
A few evaluations have been made, mostly comparing gas-phase reactions to their adsorbed counterpart to demonstrate a lowering of the barrier. This is already promising, since Density Functional Theory (DFT) often used in these complex systems has been shown to give results opposed to observation. In one example, the reaction appears to be more difficult on the catalyst surface. \cite{hog2}  In other DFT work, as the pseudo-potential improves or the number of electrons approaches the all-electron system, the barrier, initially far below that measured, continues to decrease. \cite{kroes} These tendencies for DFT results make such an approach unsatisfactory for the bench-marking of barriers. DFT can be used to explore surrounding potential energy surfaces reliably with {\it ab initio} or experimental input. \cite{diaz}

Quantum Monte Carlo (QMC) calculations are the method of choice. A pre-requisite for applying QMC is access to a super computer. In early work, a fixed geometry was used, since optimisation was prohibitively long for large systems. No direct experimental geometry for transition state (TS) species is available and use of reaction intermediate structure may not be adequate. DFT TS structure input was previously used but geometries may be inaccurate.

The CASINO code is used, its forces algorithm \cite{nemec} gave a molecular active site Slater type orbital TS, expanded in plane-waves like the periodic solid (extending our previous work \cite{hog1}).

The all-electron system is prohibitively large and transferrable atomic core electrons must be replaced by effective-core potentials, or pseudo-potentials. Note, however, that these core-electrons account for the majority of the total correlation energy \cite{clem}. The core-size must be chosen small enough to leave all electrons influenced by interactions among the valence electrons (treated explicitly). The core must be obtained at the Dirac-Fock level (to include scalar relativistic effects, notable on inner electrons) and include non-linear core correction that was fist evidenced by Louie and which needs to be included in order that the pseudo-potential can account for the total core-valence interaction \cite{nlcc}. After these choices, non-local effects can require special treatment (see below). With these provisos, electron correlation contributions approach the all-electron limit.

For CASINO, the PP must be norm-conserving and a {\bf local channel} in the (spherical harmonic) L-expansion must be chosen.

Interested readers can refer to the extensive study, which shows non-locality can cause havoc, if the core is chosen too large, the local channel badly defined and above all for the 3d electrons \cite{kdd1}. The 4d and 5d are 10 times lower in variance, ascribed to the lower dispersion of the electron distributions, for increasing Z$_{eff}$.


{\bf Quantum Monte Carlo calculations are carried out in two steps.}

The variational (VMC) step is used to provide a trial wave-function as the product of a Slater determinant and so-called Jastrow factor of explicit correlation variables. The latter is expanded in polynomial form and the coefficients optimised during the variation step, preferably with respect to ground-state energy.

Afterwards, this wave-function is used to generate a population of real-space (co-ordinate space) configurations that are propagated in imaginary time during the second, diffusion step (DMC). DMC is carried out in the fixed-node approximation, which uses the nodes from the input trial wave-function. An updated overview of the VMC and DMC methods is given in \cite{cas2020}. These trial wave-functions can be optimised with a complex Jastrow factor, \cite{garnet}  because they potentially provide the input with exact nodes. This improves the single Slater determinant describing a ground-state from the DFT orbitals for heterogeneous systems. DFT nodes may well be poor. Multi-determinant input is inaccessible for large systems at present (however, multi-determinant wave-functions and even large CI trial wave-functions have been shown to improve the wave-function nodes for diatomic molecules \cite{kdd1}). A compromise has been applied in this work by embedding a molecular active site at the MRCI level into the periodic solid (in this case, platinum). The whole wave-function is then expanded in plane waves.

\section{The Quantum Monte Carlo Method: specifications.}

\hskip4mm This work evaluates reaction barrier heights. The transition-state geometry is optimised using QMC. The experimental asymptotic equilibrium CO and water geometry (8 \AA \hskip1mm from the defect-free surface) is used. Subtracting the corresponding energies eliminates most of the non-locality and fixed node non-systematic error \cite{kdd1}.

The present study simulates heterogeneous catalysis that enhances bond dissociation. This step is frequently the initial (and often limiting) step of an industrial reaction (see, for example \cite{kroes}). Bond dissociation is difficult to describe using most quantum theory approaches, even for isolated diatomic molecules. Taking electron correlation into account does lead to the prediction of the observed products.

Therefore, heterogeneous catalysis requires an approach correctly giving almost all electron correlation (the related energy varies as bonds dissociate), in addition to the interactions involved.
It is well-known that Hartree-Fock methods fail to describe bond dissociation correctly. This is so, even for homo-nuclear diatomics. Results closer to observed products need lengthy extensive Configuration Interaction (CI) to cater for electron correlation. DFT methods present a rapid computational alternative, with some correlation which performs better in the dissociation limit when certain functionals are used.

The Perdew, Burke, Ernzerhof (PBE) functional is used since it previously gave accurate lattice parameter values. It gives a reasonable gas phase barrier for ammonia synthesis, however, on Pt(111) limiting the relaxed compact cubic structure, the barrier increased significantly, as opposed to observed catalytic effects \cite{honk}.

QMC which includes correlation explicitly is required. The trial wave-function must behave as correctly as possible, in particular close to dissociation. A high-level wave-function embedded in periodic  Kohn-Sham PBE plane-waves is a suitable starting point for trial wave-functions in Quantum Monte Carlo work on transition metal systems (see below).

In QMC, electron correlation is uniquely well accounted-for. The corresponding energy contribution varies dramatically during adsorption and reaction, therefore it must be determined exactly. A set of DFT benchmarks by Thakkar \cite{thakk} for evaluating electron correlation energy highlights the poor performance of 11 much used DFT functionals. A comparative study of Diffusion Monte Carlo (DMC) and DFT-MRCI (DFT-Multi-reference configuration interaction) methods for excitation energies tests similar cases to the present rate-limiting activation barriers (in system size and electron re-arrangement): percentage errors in the total excitation energies are 3 \% for DFT-MRCI and only 0.4 \% for DMC. (See Lester \cite{les}).

The Quantum Monte Carlo approach uses statistical physics over a large population, comprising sets of instantaneous particle positions in co-ordinate space. They are often called 'walkers' (by analogy with the one-dimensional random-walk, random numbers actually serve to initialise the 'walkers' from the trial wave-function that defines the initial electron density population). After equilibration, for numerous data-points N, high accuracy is obtained with the error decreasing as 1/$\sqrt{N}$.  Trial wave-function quality is carefully optimised.and finite size effects catered for.  Solid-state QMC can be made to scale slowly with system size (n electrons scale as n$^3$), expanding the plane-wave basis in cubic splines (blips) \cite{alfe}.
The CASINO code is used, which is well-suited to periodic solids.

The purpose of the work reported here is to study the stabler of two possible catalytic reaction paths for adsorbed CO+H$_2$O, distinguished via their specific TS.

The statistical error (0.2 kcal/mol) within which the calculated reaction barrier heights are located, must be taken as including specific non-systematic contributions (due to non-locality of the pseudo-potential and poor nodes of the trial wave-function).

In our previous work, two distinct mechanisms for CO and water co-adsorption providing hydrogen synthesis have been put forward for the reaction on Pt (111) \cite{absi}.

I-a step by step process, with rate limiting water dissociation on Pt \cite{phat}.

II-a concerted step, with CO and water co-ordinated to the metal.

This work studies the concerted mechanism (II): QMC energies of asymptotic physisorbed geometries for CO and H$_2$O are subtracted from that of the stringently optimised QMC geometry for the adsorbed Transition State.

The transition state we obtain using QMC force constants (the Hessian is defined with QMC force constants as matrix elements, then updated using Pulay's Direct Inversion in the Iterative Subspace (DIIS) algorithm adapted by Farkas \cite{fark}). The resulting TS is shown in Figure 1. It has a slightly short CO bond, since some electron density involved in beginning the bonding with the water oxygen comes from a CO anti-bonding orbital. The water molecule has a very long OH-bond (1.2 \AA), since the hydrogen is bound to Pt at the origin, whereas the other O-H is almost unaffected.

In this study, we have established its geometry, using optimisation guided by QMC force constants and precisely estimated the activation barrier height.

\section{Setting up the model system.}

\hskip4mm These systems involve stretched bonds requiring almost all the electronic correlation. For such geometries, our previous work has shown \cite{boufh} that QMC captures much more correlation energy than even PBE-DFT. QMC is therefore reliable for near-exact correlation in bond-dissociation, contrary to DFT and even the correlated Hartree-Fock based methods.
\newpage
A Jastrow factor of polynomials accounting for explicit correlation is multiplied into the Slater determinant wave-function to account for electron correlation.
For applications to this heterogeneous catalysis, a slab of the metal is constructed. This is reapeated periodically in both directions (x and y) within the solid and defining a planar (111) surface at z=0. Periodicity in the z-direction is introduced by repeating the slab after a large vacuum specing (20.8 \AA) and is limited to the half-axis 'outside' the catalyst tat includes the adsorbed molecules, with the bulk z axis retaining platinum periodicity perpendicular to the Pt(111) face. The x and y dimensions should be made much longer than the maximum bond-length and sufficiently thick for the surface layer perturbation during geometry optimisation to be attenuated at a depth within the slab to the extent that the bulk parameters apply to the final layers. The solid is a fixed experimental geometry 2 by 2 five layer platinum slab (20 atoms) This allows the close-packed alternation ABABA, exposing identical faces in the slab.

The surfaces in this model are planar, whereas it is known that metal close-packed faces re-arrange forming high Miller index ridges or meanders. \cite{pimp} Our planar-surface model uses the following states:

The reference state with the same atoms as the TS (i.e. just before the reaction): (1) The physisorbed asymptote for distant H$_2$O. The water molecule is placed 8 \AA \hskip1mm from the above slab-construct with pre-adsorbed CO. A double cell run per molecule is needed.

(2) Co-adsorbed molecules are 2 \AA \hskip1mm above the Pt(111) plane but 8 \AA \hskip1mm apart.

(3) The Transition-state (TS) geometry (see Figure 1).

\begin{figure}
\includegraphics[scale=0.5]{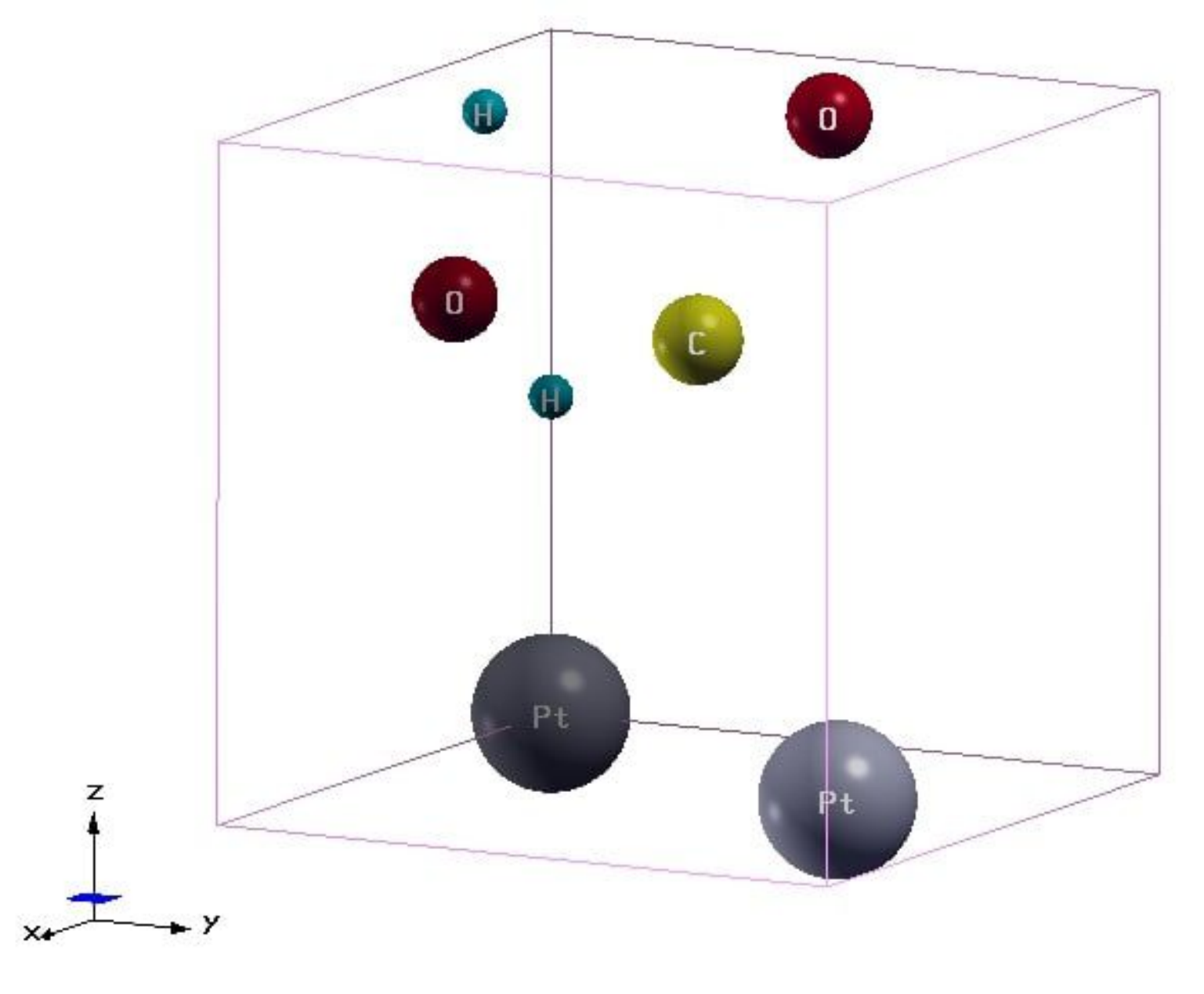}
\caption{Molecular active site (TS): the z=0 plane is the Pt(111) surface (bottom face). Pt at origin accepts one water H.  CO bond-length 1.43 \AA, Carbon above a hollow site. Water (oxygen) O--C distance 1.63 \AA.}
\end{figure}

This geometry 3 serves to initialize the QMC transition-state, placed on the slab replacing its four-Pt atom skin and adsorbate. It is then treated as input for a self-consistent plane-wave calculation to obtain the single-determinant for the TS trial wave-function. The reference energy is that corresponding to the geometry in step 1.

To limit finite-size effects, a maximal k-point grid defines a real-space super-cell. Computer memory limits its size to one much smaller than the converged DFT/plane wave grid.
In this work, we were able to test a 3 3 1 grid, assessing corrections to finite-size effect.

The DFT calculations converge only at 16 16 1 or higher. They are useful as control variate in twist-averaging. It corrects for finite size error which cancels between asymptotic and Transition State geometries leaving about 1/10 of the initial total error. Runs on a 2 2 1 grid reduced finite size-error by a factor 25 (127 for 3 3 1) compared to a gamma-point calculation.

 The pseudo-potential must allow inclusion of semi-core electrons in the valence is certainly necessary when dissociating molecules involving the transition metal atoms. It also improves test results for platinum. The copper 3d$^{10}$ shell is dense in the core-region but platinum has its 5d$^9$ 6s$^1$ ground-state shell on average further from the core. Pt is thus less prone to difficulties defining pseudo-potentials. Semi-core electrons also appear to have much less influence on metallic systems in which the number of metal atoms is conserved and geometry very similar (almost bystanders). Such is the case when comparing asymptote and QMC optimised TS geometries for reaction barriers. Co-ordination of the TS to the metal surface occurs but the distance between atoms is about 40 \% more than the equilibrium bond--length and so the role of semi-core electrons is minor. A Z=60 core for Pt is validated (see below).

\section{Trial wave-function and pseudo-potential (PP).}

\hskip5mm The trial wave-function was evaluated for a molecular active site using MOLPRO \cite{molpro}. This 'molecule' comprises a Pt-atom at the origin and a second Pt to define an apex of the triangle motif in Pt(111). The CO is adsorbed above the centroid. For Pt-atoms, we used the Dirac-Fock Z=60 effective-core potential for Pt, denoted ECP60MDF \cite{fpseu} leaving 18 valence electrons per atom. For this molecular active site, the MOLPRO AVTZ contracted Gaussian basis is used for spd orbitals.
The initial Hartree-Fock is based on the fact that the system has a closed shell of electrons (all are paired). The Restricted HF (RHF) code is used in Molpro. A number of post-Hartree-Fock methods were tested, not all of which were easy to converge for the TS-geometry. The TS-geometry used was converged for MRCI. All energies in Hartree.
\begin{itemize}

\item{Input Geometry (Bohr):  TS geometry for molecular active site.}

  Pt,   0.0000     0.0000     0.0000

  Pt,   2.6137     4.5271     0.0000

   C,   2.6137     3.0179     4.3000

   O,   2.6137     4.4344     6.5864

   O,   1.9000     0.0000     4.3000

   H,   0.0000     0.0000     3.0500

   H,   1.3155     0.0000     6.0126

\item{Perturbation}: MP2 Moller-Plesset second order.

  E$_{RHF}$ =Reference energy         -424.65248

  MP2 singlet pair energy              -0.86229

  MP2 triplet pair energy              -0.64981

  MP2 correlation energy               -1.51210

  MP2 total energy                   -426.16459

\item{Multi-Reference CISD}

Configuration Interaction, Singles and Doubles (internally contracted)

 E$_{RHF}$ =Reference energy         -424.65248

Cluster corrected energies          -425.75082 (Pople, relaxed reference)

\item{CISD}

Did not meet convergence criterion.

\item{FCI}

Did not meet convergence criterion.

Carried out with NECI Full CI    -426.28077

\end{itemize}
\hskip5mm First, a Multi-Reference CI is carried out for the transition-state (TS) and asymptote geometries. The transition state we obtain using QMC force constants (the Hessian is defined with QMC force constants as matrix elements, then updated using Pulay's Direct Inversion in the Iterative Subspace (DIIS) algorithm adapted by Farkas \cite{fark}). The resulting TS is shown in Figure 1. It has a slightly short CO bond, since some electron density involved in beginning the bonding with the water oxygen comes from a CO anti-bonding orbital. The water molecule has a very long OH-bond (1.2 \AA), since the hydrogen is bound to Pt at the origin, whereas the other O-H is almost unaffected.

The molecular wave-function is then evaluated to Full-CI, with NECI \cite{neci}and the top 20 determinants selected. It is embedded into the 5 unit-cell thick Pt-slab representing the lattice and expanded in plane-waves. This Pt slab comprising 4 atoms in each of 5 layers is defined as face-centred cubic, exposing 2 identical compact (111) faces i.e ABABA.

The molecular active-site is defined the wave-function obtained from MOLPRO and NECI, as explained above.  This is embedded in the Pt slab using the method of \cite{bens} i.e. Green's function for electronic potential additivity and the whole system wave-function is expanded in plane-waves expressed as B-splines \cite{splin}.

The 5-layer Pt-ABABA slab top 3 layers are spaced according to experimental measurement. The last two derive from the exact bulk a = 3.912 \AA. In plane Pt(111) atom-spacing is always $ a \over \sqrt{2}$ and the last layers are $ a \over \sqrt{3}$ apart to ensure continuity with the bulk. Periodicity for {\bf in plane (x,y)} directions is physical and on the z-axis the slabs are repeated with a vacuum spacing of 20.8 \AA. The super cell used has 1512 electrons, because it is expanded in real space over a 2 by 2 by 1 grid, multiplying the cell and also decreasing finite size effects by spacing the adsorbed molecules by a distance L that is much larger than the overall molecular diameter: Here we have L=10.5 au and the molecular diameter is 3.6 au.


Previous work used Troullier Martins norm conserving pseudo-potentials (TMPP), which were shown to be inadequate in terms of accuracy for the test reaction below, giving PtH.

The present work therefore converts the Effective Core Potential (ECP) used in Molpro for the active site molecular wave-function to a norm conserving form which has the same local channel (l=1) as ECP60MDF. Care must be taken to define this correctly, since the TMPP generally opts for the highest l-value as local channel (i.e. l=3 for the Pt TMPP). A study of the best practices for choice of local channel and the non-locality defect can be found detailed generally in \cite{kdd1} and tested specifically for the Pt-atom in \cite{Raj1}. Ignoring the non-locality problem leads to VMC optimisations that behave differently according to the quantity minimised. The CASINO 'varmin' algorithm (that minimises variance) can give artefactal Jastrows that take DMC to significantly sub-variational energies. This, is not in itself absurd, if limited, however use of Casula T-moves becomes compulsory and because of the truncated 'non-local' integral expansion it can still be too far below. This is not happening since tests of various PPs for e.g Pt in simple compounds (like Pt-H, for which we have reference CI wave-functions from Molpro, so the reaction Pt$_2$ + H$_2$= 2PtH also tests the adequacy of the options at DMC (and, most importantly the PP).

The integrals over PP (including the Casula T-move correction for non-locality) are numerical with truncated series (some loss of accuracy).

The TMPP is relatively hard, conferring low variance on the wave-function. It was halved by careful optimisation of the Jastrow factor (see below).

One cause of this poor initial condition is that pseudo-potentials are difficult to construct when atoms possess d-electrons that penetrate the region of space occupied by the core whilst also being diffuse and therefore of significant electron density in the outer region of the atom where the valence electron density is high. For copper, the 3d electrons are explicitly valence. Variance is much lower for Pt slabs. A Pt core is quite free of d-density \cite{hog2} . Each Pt atom possesses the electronic ground-state (5d$^9$ 6s$^1$).  It is not very well-described by 10 active electrons (9 in 5d orbitals with some density in the core-region). This is demonstrated using a Troullier-Martins pseudo-potential which provided a QMC trial wave-function from PBE Kohn-Sham orbitals, compact and with low variance per electron of only 0.025 au (an order of magnitude less than that for copper treated similarly) and yet failed on the test addition reactions of platinum, involving hydrogen:
\vskip1mm
Pt + H$_2$ (giving PtH$_2$ (a) and Pt$_2$ + H$_2$ giving 2 Pt-H (b).
\vskip1mm
The QMC reaction energy for (b), that has been shown to proceed via a squarish/rectangular TS, is 5-6 kcal/mol above that from reference (MRSDCI) calculations \cite{balb}.

For Pt$_2$ + H$_2$  a binding energy of 35.2 kcal/mol is predicted by QMC instead of the 30.7 kcal/mol obtained from MRSDCI benchmarks. This is ascribed to residual local approximation errors.
We designed a hard pseudo-potential (with 18 electrons treated explicitly) which is based on the MDF60 PP of the Stuttgart group 15 (which treats as valence 9*5d 1*6s and 8*4f, most notably retaining 6 of the 4f electrons in core) and found it allowed us to accurately simulate the test reaction molecules. The QMC binding energy was 31.1 kcal/mol (c.f. 30.7) 14 with a standard error of 0.5 kcal/mol.

\subsection{Limiting possible systematic error in QMC.}

\subsubsection{Time-step. Check for bias.}

\hskip5mm A short time-step is necessary for such heavy atoms  An initial time-step of 0.001 au was used for the Diffusion Monte Carlo stage of the calculation, giving almost exact ground state combined system properties. No time-step bias was observed by extrapolating to zero. The zero time-step limit is valid for small finite time-steps used here.

\subsubsection{Fixed or released nodes.}

\hskip5mm Deficient fixed nodes in the input determinants may result from the Kohn-Sham orbitals, limitation in the CI or the use of pseudo-potentials..

Fixed-nodes are relaxed when using a complex Jastrow factor \cite{garnet} greatly improving the input. This 'exact' approach was compared with the single real generic Jastrow approach described below,  which was eventually selected for the numerous statistics-accumulation DMC production runs.

\subsubsection{Finite size effects.}

\hskip4mm Adsorbed system work often uses a clean metal surface reference. This implies that extensively delocalised states are involved, which formally describe conduction bands of electrons. In DFT work on metals using software for periodic solids with plane-waves, the first Brillouin zone shows that these states and the surface energy converge slowly with the size of k-point grid. The present state of the art QMC for solid state applications use a modest size of grid because it must be 'unfolded' for QMC. This means the number of particles is repeated the same number of times as there are k-points. Often, therefore, the QMC calculation is limited in practice to a grid that is small with respect to convergence. In practise, the wave-function input data must fit the available computer node-memory.

Here a 2 2 1 grid where the 18-electron Pt-atoms in the cell are repeated four times. There are 5 Pt-layers and 4 per layer, givin 1440 electrons per supercell.  This improves the energy error in surface formation for clean metals, as shown by previous work on copper 2 by over a factor 25, compared with the usual single k-point but a short-fall with respect to convergence remains. This phenomenon is referred to as the finite size effect. Here, the activation barrier calculation refers to CO and H$_2$O physisorbed on this 2 2 1 Pt(111) grid, with 4*18 more electrons, giving a total of 1512 electrons, instead of clean Pt. Including physisorbed reactants in the reference limits the effect of delocalised conduction band states. Those remaining with 2-D symmetry are less prone to finite size effects. This strategy represents a definite progress overall error remaining is less than 2mHa.  These errors are due essentially to 3-D delocalised states involved. Finite-size error may be greatly reduced by twist-averaging and some further analytical corrections for the single-particle asymptotes. \cite{asy} At each twist of the 25 used here, re-equilibration is followed by collecting 100 000 data-points. Converged DFT energy (at each twist) is a control variate \cite{con}.
These key problems are alleviated when comparing very similar geometries comprising the same atoms (with wave-functions describing quite similar states).


\section{Variation Monte Carlo:}

A preliminary Variational Monte Carlo (VMC) calculation is carried out in order to generate several thousand configurations (instantaneous points in electron co-ordinates). VMC is driven by energy minimisation. The so called {\it local energy}:  ${H \psi } \over \psi$ is evaluated, including kinetic energy terms that are smoother and have lower variance in exponentially decaying bases, as shown in our work on wave-function quality \cite{wfq,tlse}. A Jastrow factor including electron pair, electron-nucleus and three-body (two-electron and nucleus) is defined. This Jastrow factor \cite{jas} is carefully optimised (essential work, taking up to 5 \% of the total time). This factor uses a polynomial expansion in the variables of explicit correlation. With complex Jastrows, the wave-function nodes shift, otherwise not.

The product of a Jastrow factor with a Slater determinant gives the trial wave-function. In this work, the linear Jastrow optimisation is used directly on the trial wave-functions.
\subsection{Generic Jastrow factor}
\hskip5mm Defining a Slater-Jastrow wave-function as the product of $\Psi_S $, a Slater determinnt and an explicit correlation Jastrow factor J(R), to be detailed below:
\begin{equation}
\Psi{\bf(R)} = e^{J(R)} \Psi_S (\bf{R}) = e^{J(R)} D_u (R_u) \, D_d (R_d)
\end{equation}

The determinants $D_u (R_u)$ and $D_d (R_d)$ treat 'up' and 'down' spin separately if necessary. A new approach \cite{con} but which generates a huge parameter set was tested for this platinum slab model. A generic Jastrow factor (using Common Algebraic Specification Language (CASL)), with the e-e, e-n and e-e-n polynomials expanded to order 9 with cut off radius of 10 au.

Initialising the generic Jastrow factor give an indication of the cut-off radii (values) and its structure.
Truncation may be used to a finite range with a factor:
\begin{equation}
t(R) = (r-L)^C \Theta_H(r-L)
\end{equation}
The wave-function must be continuous and possess at least two continuous derivatives (i.e. C=3). $\Theta_H$ is the Heaviside function, L the range, e.g. diameter of the Wigner Seitz cell.

\begin{equation}
J({\bf R}) = \sum_{i<j}^N u_{P_{ij}} (r_{ij}) \, + \, \sum_{i}^N \, \sum_{I}^{N_{n}} \chi_{S_{iI}} (r_{iI}) \, + \, \sum_{i<j}^N \sum_{I}^{N_{n}} f_{T_{ijI}} (r_{iI}, r_{jI}, r_{ij})
\end{equation}
$u, \chi $ and $f$ are parameterised (polynomial) functions in the inter-particle distances. The symbols P, S, T denote the channel indices. The u term caters for inter-electron contributions over $r_{ij}$ and the $\chi$ term for electron-nuclear terms over $r_{iI}$ . They are both constructed to obey Kato's cusp conditions. The $f$ term is the three-particle electron pair and nucleus contribution (over $r_{ij}$ and two $r_{iI}$ values. This 3-body $f$ term cannot be neglected.

An example of the initial generic Jastrow factors we have used to begin optimisation is in Appendix 2. The fully optimised TS and Asymptote generic Jastrow factors are given in Supporting Information.

This Jastrow factor is fully optimised during the VMC step, first by using the mean-average deviation that reduces variance in the distribution whereby the configurations (instantaneous particle positions in co-ordinate space) represent electronic density.

This distribution is updated until the final stages of VMC where fine-tuning of the Jastrow factor is accomplished by energy minimisation over 15 cycles. Our fully-optimised generic Jactrow factor is given in the online repository \cite{bcs}.

The resulting Slater-Jastrow wave-function is used to initialise the configurations for Diffusion Monte Carlo (DMC). This gives prohibitively long DMC cycles, because three parameter-sets are obtaained for each of the 25 atoms in the super-cell, treated individually. It does, however, seem to approach the variational energy minimum for the system defined (-448 Ha, with the PP from ECP60MDF with l=1 as local channel). In order to cut the DMC cycle duration, individual parameter-sets are maintained only for the atoms in molecular sepecies (CO and water). The platinum atoms are grouped according to rules for surface atoms and bulk atoms. These rules allow us to use a single parameter set for each of the inter-particle terms for each group. The price to pay is that we use a sub-optimal Jastrow throughout, reaching a similar minimum energy, which we know to be somewhat above the variational limit. It is therefore of paramount important to conserve the same rules for definition of the generic Jastrow-factor in each of the limiting geometries (TS and asymptote). There is little risk in violating the correlation space spanned by the parameter-sets used in the present work, because the same atoms are involves, with some or negligible surface interaction for the TS and asymptote, respectively.

After the generic Jastrow optimisation has reached a stable minimum, a final VMC calculation generates the initial configurations required for the Diffusion Monte Carlo step (DMC); 10-20 per core. The previous VMC steps must generate at least as many configurations.


\section{Diffusion Monte Carlo:}

\hskip4mm In the DMC method the ground-state component of the trial wave function is projected out by solving the Schr{\"o}dinger equation (SWE) in imaginary time.

This is accomplished by noting that the imaginary-time SWE is a diffusion equation in the 3N-dimensional space of electron coordinates, with the potential energy acting as a source/sink term.

The time-dependent Schr{\"o}dinger equation may be written as:
\begin{equation}
i \, \frac{d}{dt} \Psi(r, t)  = - \frac{1}{2}  \nabla^2  \Psi( {\bf{r}}, t) \, + \, V({\bf{r}}) \Psi({\bf{r}}, t)
\end{equation}
where $\, \nabla^2$ is the scalar Laplacian of an instantaneous electron-position vector: $\, {\bf{r}}$.
\vskip4mm

Transforming $t$ into the pure-imaginary time $it$ and defining the Hamiltonian for stationary states as:
\begin{equation}
\mathcal{H}  = - \frac{1}{2}   \nabla^2  \, + \, V({\bf{r}})
\end{equation}
Leads to a diffusion equation, in which excited state contributions carry an exponentially decreasing factor and fade out as the 'time' variable $ \tau =it$ is increased in very numerous small steps. The best-estimate of the ground-state energy eigenvalue (E$_0$) is adjusted.  Note that the exact value gives a zero derivative
\begin{equation}
- \, \frac{d}{d \tau} \Psi( {\bf{r}}, t)  = (\mathcal{H} - E_0 ) \, \Psi({\bf{r}}, t)
\end{equation}

The imaginary-time SWE can therefore be solved by combining diffusion and branching or dying processes.  Introducing importance sampling, using the trial wave function transforms the problem into one involving drift as well as diffusion but greatly reduces the population fluctuations due to the branching/dying process. The Fermionic wave function antisymmetry is maintained by constraining the nodal surface to equal that of the trial wave function.

The statistical error bar $\sigma$ on the QMC total energy must be small compared with the energy difference to be resolved.  Assuming the cost of the equilibration phase of a QMC calculation is negligible, $\sigma$ falls off as 1/$\sqrt{T}$, where T is the cpu-time in core-hours.

To sum up, the accuracy for a given wall-time depends on the quality of the trial wave-function and overall accuracy essentially increases with the square-root of the number of data-points collected. Production runs used 4096 cores (hybrid MPI/MP parallelization gives 4 threads/core). For each twist up to 10 runs for each geometry (water-dissociation TS, concerted TS and asymptote) were needed, with a maximum check-point interval of 20h.

The DMC algorithm  proceeds using Casula T-moves and a time-step of 0.005 au. These are continued until error in standard error is 0.1, i.e. for 27 500 data points on average for 16 twists (the wave-function is complex since the structure has no inversion center).

\section{Results.}
\hskip4mm  The water-gas shift reaction on Pt(111) is a source of  sustainable H$_2$ production. This reaction of carbon monoxide and water is practically unknown in the gas phase. We use QMC to describe CO+H$_2$O adsorbed on a Pt(111) surface. The surface has 2-D periodicity. Molecules interact with both the surface and each other. A slab construct, limited by relaxed Pt(111) (reaching the experimental bulk geometry in 4 layers) is used. Carbon monoxide (CO) adsorption is considered. Carbon monoxide gas is rather inert and the carbon atom even carries a small partial negative charge.

Upon adsorption (after physisorption that little alters CO), charge transfer to a Pt-surface restores the typical reactive carbonyl species, with a positively charged carbon that is the site of nucleophilic attack  (with a partial charge 0.098 e). The strong bonding within this molecule when in the gas phase is weakened by interactions with the surface, making the molecule easier to attack by water etc. The carbon monoxide molecule is thereby polarized by its adsorption on the surface that results it becoming reactive toward nucleophiles.

Investigation of this phenomenon has been carried out using infrared reactor cells (in situ FTIR, \cite{baz}) at a number of solid surfaces. Here, for Pt(111), it is complemented by a QMC study of the reaction with water. Removal of toxic carbon monoxide molecules is a typical de-pollution reaction, of interest in catalytic exhausts. Furthermore, the reaction with water is of industrial importance in producing clean fuels (hydrogen gas) in a sustainable process.
In earlier work \cite{hog2,baz},it was shown that the stretch frequency of carbon monoxide makes it a highly sensitive probe to surface interactions. This frequency is a measure of adsorption. The carbon monoxide group also acquires the {\bf carbonyl} polarization, i.e. is reversed to become the site of nucleophilic attack.

{\bf CO and water:}

 These co-adsorbed molecules react, as shown by preliminary investigations using {\it ab initio} DFT (PBE functional) studies, with a plane-wave basis. The reaction products are CO$_2$ and gaseous hydrogen (fuel restoring water as combustion product). CO adsorbed on Pt (111) was optimized using QMC force constants and found to be 118 pm, only slightly longer than the gas-phase value of 113 pm which is partly due to the interaction with water depleting electron density of a CO anti-bonding orbital). This CO can be studied by FTIR and shown to possess a partial positive charge on carbon, suitable for nucleophilic attack. Density analysis of QMC results shows the partial charge is +0.098 e at a Pt-C distance of 1.82 \AA.
Water dissociation concerted with oxygen beginning to link with the CO carbon is rate limiting. Dissociation preceding reaction is less favorable but was the reaction channel initially studied.
In Figure 1, (see the above {\bf model} section) we show the present QMC geometry of the TS. The water molecule with the leaving H$^{\delta+}$ binds with a surface Pt-atom, whilst the remaining O-H is forms a V-shaped H--O$^{\delta-}$--C$^{\delta+}$=O intermediate. The carbon-monoxide carbon atom and water oxygen are 2.27 \AA \hskip1mm above the Pt(111) face. The structure was initialized from recent DFT work \cite{faj}. The resulting QMC optimized geometry determines three Pt-O tripod distances as 2.04 \AA. The O-H bond-lengths are 0.98 \AA \hskip1mm and 1.16 \AA.

The O-C linkage forming is still 1.63 \AA \hskip1mm long (in this TS).
The four-Pt atom {\bf skin} lozenge, are at the apices of a trigonal (i.e. hexagonal) close-packed cell with Pt-Pt distance 2.783 \AA.

Estimated barrier heights in kcal/mol:

This comparison between DMC simulations of the TS and asymptote geometries (structures 3 and 1 of the model section)  after averaging over 25 stochastically generated twists for each geometry using the CASINO/pwscf interface for reactions I and II:
Some mechanistic studies of the water gas-shift reaction \cite{phat} have suggested water dissociation is rate limiting and followed by attack of CO (I). This scenario was studied in the initial stages of this work, however, a concerted TS (II) illustrated in Figure 1 has been found in this work and was previously postulated \cite{mukin}.
\vskip4mm
I-water dissociation at Pt(111): QMC barrier of 17.74 (standard error is 0.3 kcal/mol).

This is in good agreement with PW91 USPP values: 17.99 kcal/mol.
(Recent DFT work from \cite{phat}).
N.B. gas-phase water dissociation has a barrier of 117.5 kcal/mol.

\vskip4mm

II-water dissociation concerted with the start of nucleophilic attack of CO at Pt(111):
QMC Study based on the TS of Figure 1 above, in the Model section (with the carbonyl polarization of adsorbed CO, i.e. partial positive charge on carbon unlike CO (gas)).

The final QMC barrier of 17.0 with a standard error of 0.2 kcal/mol. This QMC value could be taken as a lower bound for the activation barrier, limited by the activated complex. The actual system will depend on temperature, surface re-arrangement and possible defects. Agreement is perfect with this model: apparent activation energy measured is 17.05 kcal/mol.

\vskip4mm
\section{Perspectives and conclusions.}

\hskip4mm This Quantum Monte Carlo determination of the Transition state structure and reaction barrier height for water gas shift provides new insight into its use for producing hydrogen as a sustainable energy source, with an industrial catalyst. From this work, the concerted mechanism appears favored.

The catalyst studied is platinum, acting via its (111) close-packed surface. A small molecular structure, described by Slater type atomic orbitals is optimized using QMC forces and the CASINO software. This geometry is used to initialize plane-wave DFT wave-function evaluation providing Kohn-Sham PBE orbitals to define the trial QMC wave-function completed by a complex Jastrow factor. This embedding procedure is validated for some test reactions of hydrogen on platinum clusters. The Pt(111) slab used here was also tested.
The barrier height is therefore not expected to suffer from systematic errors and the statistical (standard) error of 0.2 kcal/mol is realistic, and comparable to that of measured appatent activation energy.

Perspectives for extending this strategy come from full-CI input for molecules available in the Alavi group. At the time of writing, this code provides QMC force constants for all-electron calculations on molecules. The method needs to be tested for effective core potentials (ECP) and the geometry embedding for plane-wave input needs to be benchmarked. This is the aim of one of our current research projects.


25 offset grids paving the first Brillouin zone allow for twist-averaging and reduction of the standard error. Mechanistic and barrier height results are given here, pending a detailed methodology paper, currently in preparation.

The present work obtains better than chemical accuracy (i.e. energy to within 1 kcal/mol) for the rate-limiting step and corresponds well to measured apparent activation energy (17.05 kcal/mol c.f the QMC value we get: 17.0 $\pm$ 0.2 kcal/mol). The molecular active site has a Full Configuration Interaction (FCI) wave-function and this is embedded in a periodic Density Functional Theory (DFT)/plane-wave lattice.

We refer to CO adsorbed above the centroid of an equilateral triangle with Pt-atoms as apices (defined using translational symmetry from a single unique Pt-atom) in the Pt(111) plane. This gives three C-Pt linkages.


On the Irene CEA supercomputer near Paris, 25 DMC steps take 45 mins on 4032 cores.

\hskip8mm The data-set supplied for the given k-points in the first Brillouin zone defined for the Transition-State and referred to the asymptote give the tabulated values below:
A test on time-step bias yields 30 micro Hartree for the energy difference, or 0.0019 kcal/mol and will be neglected. Tabulated values are for the DMC population target weight.
\vskip2mm
\textcolor{blue} {Summary of data from: 10.6084/m9.figshare.10293194}

{\bf Table I:Activation energy evaluation for the online data-sets from this work}
\vskip1mm
 \begin{tabular}{|c|c|c|}
  \hline
 Structure & Transition-state  & Asymptote   \\
 \hline
 E$_{tot}$ (Ha) & -549.63218 & -549.65924  \\
\hline
Variance au/(kcal/mol)$^2$ & 0.604/0.2377 & 0.528/0.2081  \\
\hline
\end{tabular}
\vskip2mm
Activation barrier: 0.2706 au or 16.98 (17) $\pm$ 0.67 kcal/mol (this standard error here is obtained from data-analysis of the distribution at the gamma point. It is reduced by almost a factor 4 by twist-averaging i.e. combining independent runs from 16 offset grids in the first Brillouin zone). The final QMC activation barrier is thus 17.0 $\pm$ 0.2 kcal/mol.
\vskip2mm

{\bf Acknowledgements.}

The data for TS structure uses CASINO QMC forces code and a contact molecule, comprising CO+H$_2$O and a monolayer of Pt(111) optimized using QMC and placed on the slab at the experimental spacing.
The Partnership for advanced computing in Europe (PRACE) backed this project.
The QMC calculations were made possible by allocation of supercomputer resources to project 2018184349: 51.6 Million core-hours on the Irene supercomputer (CEA, Bruy\`eres-le-Ch\^atel), near Paris, France.

We are also grateful for access to Finland's National Computer resources csc.fi (CSC – It Center for Science) for some computational capacity (all Molpro runs).

PEH thanks Pablo Lopez-Rios for supplying NECI and for helpful discussions.

\newpage

{\underbar {Appendix 1.} }

Supercell 2 2 1 for Pt lattice parameter 3.912 \AA \hskip2mm (7.3928 bohr) exposing Pt(111) +CO + H$_2$O (TS;

 asymptote translate water by 12.5 Bohr on z)

    1  78   1     0.0000     0.0000     0.0000     0.0000     0.0000     0.0000

    2  78   1     0.5000     0.0000     0.0000     5.2275     0.0000     0.0000

    3  78   1     0.0000     0.5000     0.0000    -2.6137     4.5271     0.0000

    4  78   1     0.5000     0.5000     0.0000     2.6137     4.5271     0.0000

    5  78   1     0.3333     0.1667    -0.1073     2.6137     1.5090    -4.2247

    6  78   1     0.8333     0.1667    -0.1073     7.8412     1.5090    -4.2247

    7  78   1     0.3333     0.6667    -0.1073     0.0000     6.0362    -4.2247

    8  78   1     0.8333     0.6667    -0.1073     5.2275     6.0362    -4.2247

    9  78   1     0.1667     0.3333    -0.2155     0.0000     3.0181    -8.4843

   10  78   1     0.6667     0.3333    -0.2155     5.2275     3.0181    -8.4843

   11  78   1     0.1667     0.8333    -0.2155    -2.6137     7.5452    -8.4843

   12  78   1     0.6667     0.8333    -0.2155     2.6137     7.5452    -8.4843

   13  78   1     0.0000     0.0000    -0.3239     0.0000     0.0000   -12.7520

   14  78   1     0.5000     0.0000    -0.3239     5.2275     0.0000   -12.7520

   15  78   1     0.0000     0.5000    -0.3239    -2.6137     4.5271   -12.7520

   16  78   1     0.5000     0.5000    -0.3239     2.6137     4.5271   -12.7520

   17  78   1     0.0000     0.0000    -0.4323     0.0000     0.0000   -17.0197

   18  78   1     0.5000     0.0000    -0.4323     5.2275     0.0000   -17.0197

   19  78   1     0.0000     0.5000    -0.4323    -2.6137     4.5271   -17.0197

   20  78   1     0.5000     0.5000    -0.4323     2.6137     4.5271   -17.0197

   21   6   2     0.4167     0.3333     0.1092     2.6137     3.0179     4.3000

   22   8   3     0.4949     0.4898     0.1673     2.6137     4.4344     6.5864

   23   8   3     0.1817     0.0000     0.1092     1.9000     0.0000     4.3000

   24   1   4     0.0000     0.0000     0.0775     0.0000     0.0000     3.0500

   25   1   4     0.1258     0.0000     0.1527     1.3155     0.0000     6.0126

   26  78   1     0.0000     1.0000     0.0000    -5.2275     9.0543     0.0000

   27  78   1     0.5000     1.0000     0.0000    -0.0000     9.0543     0.0000

   28  78   1     0.0000     1.5000     0.0000    -7.8412    13.5814     0.0000

   29  78   1     0.5000     1.5000     0.0000    -2.6138    13.5814     0.0000

   30  78   1     0.3333     1.1667    -0.1073    -2.6138    10.5633    -4.2247

   31  78   1     0.8333     1.1667    -0.1073     2.6137    10.5633    -4.2247

   32  78   1     0.3333     1.6667    -0.1073    -5.2275    15.0905    -4.2247

   33  78   1     0.8333     1.6667    -0.1073    -0.0000    15.0905    -4.2247

   34  78   1     0.1667     1.3333    -0.2155    -5.2275    12.0724    -8.4843

   35  78   1     0.6667     1.3333    -0.2155    -0.0000    12.0724    -8.4843

   36  78   1     0.1667     1.8333    -0.2155    -7.8412    16.5995    -8.4843

   37  78   1     0.6667     1.8333    -0.2155    -2.6138    16.5995    -8.4843

   38  78   1     0.0000     1.0000    -0.3239    -5.2275     9.0543   -12.7520

   39  78   1     0.5000     1.0000    -0.3239    -0.0000     9.0543   -12.7520

   40  78   1     0.0000     1.5000    -0.3239    -7.8412    13.5814   -12.7520

   41  78   1     0.5000     1.5000    -0.3239    -2.6138    13.5814   -12.7520

   42  78   1     0.0000     1.0000    -0.4323    -5.2275     9.0543   -17.0197

   43  78   1     0.5000     1.0000    -0.4323    -0.0000     9.0543   -17.0197

   44  78   1     0.0000     1.5000    -0.4323    -7.8412    13.5814   -17.0197

   45  78   1     0.5000     1.5000    -0.4323    -2.6138    13.5814   -17.0197

   46   6   2     0.4167     1.3333     0.1092    -2.6138    12.0722     4.3000

   47   8   3     0.4949     1.4898     0.1673    -2.6138    13.4887     6.5864

   48   8   3     0.1817     1.0000     0.1092    -3.3275     9.0543     4.3000

   49   1   4     0.0000     1.0000     0.0775    -5.2275     9.0543     3.0500

   50   1   4     0.1258     1.0000     0.1527    -3.9120     9.0543     6.0126

   51  78   1     1.0000     0.0000     0.0000    10.4549     0.0000     0.0000

   52  78   1     1.5000     0.0000     0.0000    15.6824     0.0000     0.0000

   53  78   1     1.0000     0.5000     0.0000     7.8412     4.5271     0.0000

   54  78   1     1.5000     0.5000     0.0000    13.0686     4.5271     0.0000

   55  78   1     1.3333     0.1667    -0.1073    13.0686     1.5090    -4.2247

   56  78   1     1.8333     0.1667    -0.1073    18.2961     1.5090    -4.2247

   57  78   1     1.3333     0.6667    -0.1073    10.4549     6.0362    -4.2247

   58  78   1     1.8333     0.6667    -0.1073    15.6824     6.0362    -4.2247

   59  78   1     1.1667     0.3333    -0.2155    10.4549     3.0181    -8.4843

   60  78   1     1.6667     0.3333    -0.2155    15.6824     3.0181    -8.4843

   61  78   1     1.1667     0.8333    -0.2155     7.8412     7.5452    -8.4843

   62  78   1     1.6667     0.8333    -0.2155    13.0686     7.5452    -8.4843

   63  78   1     1.0000     0.0000    -0.3239    10.4549     0.0000   -12.7520

   64  78   1     1.5000     0.0000    -0.3239    15.6824     0.0000   -12.7520

   65  78   1     1.0000     0.5000    -0.3239     7.8412     4.5271   -12.7520

   66  78   1     1.5000     0.5000    -0.3239    13.0686     4.5271   -12.7520

   67  78   1     1.0000     0.0000    -0.4323    10.4549     0.0000   -17.0197

   68  78   1     1.5000     0.0000    -0.4323    15.6824     0.0000   -17.0197

   69  78   1     1.0000     0.5000    -0.4323     7.8412     4.5271   -17.0197

   70  78   1     1.5000     0.5000    -0.4323    13.0686     4.5271   -17.0197

   71   6   2     1.4167     0.3333     0.1092    13.0686     3.0179     4.3000

   72   8   3     1.4949     0.4898     0.1673    13.0686     4.4344     6.5864

   73   8   3     1.1817     0.0000     0.1092    12.3549     0.0000     4.3000

   74   1   4     1.0000     0.0000     0.0775    10.4549     0.0000     3.0500

   75   1   4     1.1258     0.0000     0.1527    11.7704     0.0000     6.0126

   76  78   1     1.0000     1.0000     0.0000     5.2274     9.0543     0.0000

   77  78   1     1.5000     1.0000     0.0000    10.4549     9.0543     0.0000

   78  78   1     1.0000     1.5000     0.0000     2.6137    13.5814     0.0000

   79  78   1     1.5000     1.5000     0.0000     7.8411    13.5814     0.0000

   80  78   1     1.3333     1.1667    -0.1073     7.8411    10.5633    -4.2247

   81  78   1     1.8333     1.1667    -0.1073    13.0686    10.5633    -4.2247

   82  78   1     1.3333     1.6667    -0.1073     5.2274    15.0905    -4.2247

   83  78   1     1.8333     1.6667    -0.1073    10.4549    15.0905    -4.2247

   84  78   1     1.1667     1.3333    -0.2155     5.2274    12.0724    -8.4843

   85  78   1     1.6667     1.3333    -0.2155    10.4549    12.0724    -8.4843

   86  78   1     1.1667     1.8333    -0.2155     2.6137    16.5995    -8.4843

   87  78   1     1.6667     1.8333    -0.2155     7.8411    16.5995    -8.4843

   88  78   1     1.0000     1.0000    -0.3239     5.2274     9.0543   -12.7520

   89  78   1     1.5000     1.0000    -0.3239    10.4549     9.0543   -12.7520

   90  78   1     1.0000     1.5000    -0.3239     2.6137    13.5814   -12.7520

   91  78   1     1.5000     1.5000    -0.3239     7.8411    13.5814   -12.7520

   92  78   1     1.0000     1.0000    -0.4323     5.2274     9.0543   -17.0197

   93  78   1     1.5000     1.0000    -0.4323    10.4549     9.0543   -17.0197

   94  78   1     1.0000     1.5000    -0.4323     2.6137    13.5814   -17.0197

   95  78   1     1.5000     1.5000    -0.4323     7.8411    13.5814   -17.0197

   96   6   2     1.4167     1.3333     0.1092     7.8411    12.0722     4.3000

   97   8   3     1.4949     1.4898     0.1673     7.8411    13.4887     6.5864

   98   8   3     1.1817     1.0000     0.1092     7.1274     9.0543     4.3000

   99   1   4     1.0000     1.0000     0.0775     5.2274     9.0543     3.0500

  100   1   4     1.1258     1.0000     0.1527     6.5429     9.0543     6.0126
\newpage

{\underbar {Appendix 2.} }

Example of generic Jastrow factor input structure for VMC optimisation, as used in this work. Optimised TS and Asymptote results in SI.

The example below is a 4-layer slab, with 12 bulk atoms treated as identical. The 4 surface atoms represent one equilateral triangle (as a single polynomial) and one specific adsorption site. By default, all atoms of adsorbed molecules are treated uniquely.
\vskip1mm
JASTROW:

  TERM 1:

    Rank: [ 2, 0 ]

    e-e basis: [ Type: natural power, Order: 9 ]

    e-e cutoff:

      Type: alt polynomial

      Constants: [ C: 3 ]

      Parameters:

        Channel 1-1:

          L: [6.9899900099999996, optimizable, limits: [0.50000000000000000,
               6.9999900000000004]]

    Rules: [ default ]

    Linear parameters:

      Channel 1-1:

        c$_2$: [ 0.0000000000000000, optimizable ]

        c$_3$: [ 0.0000000000000000, optimizable ]

        c$_4$: [ 0.0000000000000000, optimizable ]

        c$_5$: [ 0.0000000000000000, optimizable ]

        c$_6$: [ 0.0000000000000000, optimizable ]

        c$_7$: [ 0.0000000000000000, optimizable ]

        c$_8$: [ 0.0000000000000000, optimizable ]

        c$_9$: [ 0.0000000000000000, optimizable ]

  TERM 2:

    Rank: [ 1, 1 ]

    e-n basis: [ Type: natural power, Order: 9 ]

    e-n cutoff:

      Type: alt polynomial

      Constants: [ C: 3 ]

      Parameters:

        Channel 1-n1:

          L: [6.8899900099999996, optimizable, limits: [0.50000000000000000,
               6.9999900000000004]]

    Rules: [ default, N, N1=N2=N3, N5=N6=N7=N8=N9=N10=N11=N12=N13=N14=N15=N16]

    Linear parameters:
      Channel 1-n1:

        c$_2$: [ 0.0000000000000000, optimizable ]

        c$_3$: [ 0.0000000000000000, optimizable ]

        c$_4$: [ 0.0000000000000000, optimizable ]

        c$_5$: [ 0.0000000000000000, optimizable ]

        c$_6$: [ 0.0000000000000000, optimizable ]

        c$_7$: [ 0.0000000000000000, optimizable ]

        c$_8$: [ 0.0000000000000000, optimizable ]

        c$_9$: [ 0.0000000000000000, optimizable ]

  TERM 3:

    Rank: [ 2, 1 ]

    e-e basis: [ Type: natural power, Order: 4 ]

    e-n basis: [ Type: natural power, Order: 4 ]

    e-n cutoff:

      Type: alt polynomial

      Constants: [ C: 3 ]

      Parameters:

        Channel 1-n1:

          L: [10.3444324177971, optimizable, limits: [0.500000000000000,
               10.4548873051022]]

    Rules: [ default, N, N1=N2=N3, N5=N6=N7=N8=N9=N10=N11=N12=N13=N14=N15=N16]

    Linear parameters:

      Channel 1-1-n1:

        c$_1$,2,2: [ 0.0E-000, optimizable ]

        c$_1$,3,2: [ 0.0E-000, optimizable ]

        c$_1$,3,3: [ 0.0E-000, optimizable ]

        c$_2$,3,1: [ 0.0E-000, optimizable ]

        c$_3$,1,1: [ 0.0E-000, optimizable ]

        c$_3$,2,1: [ 0.0E-000, optimizable ]

        c$_3$,2,2: [ 0.0E-000, optimizable ]

        c$_3$,3,1: [ 0.0E-000, optimizable ]

        c$_3$,3,2: [ 0.0E-000, optimizable ]

        c$_3$,3,3: [ 0.0E-000, optimizable ]

      Channel 1-2-n1:

        c$_1$,2,2: [ 0.0E-000, optimizable ]

        c$_1$,3,2: [ 0.0E-000, optimizable ]

        c$_1$,3,3: [ 0.0E-000, optimizable ]

        c$_2$,3,1: [ 0.0E-000, optimizable ]

        c$_3$,1,1: [ 0.0E-000, optimizable ]

        c$_3$,2,1: [ 0.0E-000, optimizable ]

        c$_3$,2,2: [ 0.0E-000, optimizable ]

        c$_3$,3,1: [ 0.0E-000, optimizable ]

        c$_3$,3,2: [ 0.0E-000, optimizable ]

        c$_3$,3,3: [ 0.0E-000, optimizable ]

\end{document}